\documentclass[aps,prc,twocolumn,floatfix,nofootinbib]{revtex4}
\usepackage{float,ulem}
\usepackage{graphicx,epsfig,epstopdf,amsmath,amssymb}
\usepackage{xcolor}

\newcommand{\be}{\begin{equation}}
\newcommand{\ee}{\end{equation}}
\newcommand{\ba}{\begin{eqnarray}}
\newcommand{\ea}{\end{eqnarray}}
\newcommand{\nn}{\nonumber\\}

\begin{document}

\title{Correspondence between momentum dependent relaxation time and field redefinition of relativistic hydrodynamic theory}
\author{Sukanya Mitra}
\email{sukanya.mitra10@gmail.com}
\affiliation{Department of Nuclear and Atomic Physics, Tata Institute of Fundamental Research, Homi Bhabha Road, Mumbai 400005, India}

\begin{abstract}
In this article a correspondence has been established between the out of equilibrium system dissipation and the thermodynamic field redefinition of the macroscopic variables through 
the momentum dependent relaxation time approximation (MDRTA) solution of relativistic transport equation. Here, it has been shown that the out of equilibrium thermodynamic fields 
are not uniquely defined and are subjected to include dissipative effects from the medium. A second order relativistic hydrodynamic 
theory has been developed including such dissipative effects. The necessary conditions for developing a hydrodynamic theory has been fulfilled, (i) the thermodynamic identities 
incorporating such redefined fields have been shown to conserve the energy-momentum tensor perfectly under MDRTA, (ii) the 
non-negativity of entropy production remains unaffected by the inclusion of such dissipative contributions in hydro fields as long as the independent transport coefficients remain positive.
\end{abstract}
\maketitle

\section{Introduction}

Relativistic dissipative hydrodynamic theory has been proved to be reasonably successful in describing the out of equilibrium dynamics of a system in the long
wavelength limit. The evolution of the relevant macroscopic quantities such as temperature and charge chemical potential is given by a set of coupled differential
equations where the out of equilibrium effects are included by certain dissipative fluxes. These fluxes are quantitatively manifested by quantities called transport 
coefficients such as viscosity and conductivities. The macroscopic thermodynamic quantities such as energy density and particle number density in these
equations are set to their equilibrium values even in the dissipative medium by imposing certain matching or fitting conditions. Recently, in a number of studies the 
inclusion of dissipative effects into these quantities have been explored \cite{Monnai:2018rgs,Tsumura:2006hnr,Tsumura:2009vm} where the out of equilibrium field 
contributions are either estimated phenomenologically or microscopically such as from the renormalization-group technique. In this article those dissipative
corrections have been estimated from gradient expansion technique of solving the relativistic transport equation using momentum dependent relaxation time approximation (MDRTA).

In recent literature, a number of studies have been carried out related to the application of momentum dependent relaxation time 
\cite{Teaney:2013gca,Kurkela:2017xis,Mitra:2020gdk,Rocha:2021zcw} in solving the relativistic transport equation and extracting the macroscopic thermodynamic 
quantities thereafter. Relaxation time approximation, originally proposed by Anderson and Witting in 1974 \cite{Relax} for relativistic transport
equation, has been used widely since then with a constant relaxation time almost everywhere. But in doing so two major issues arise that need to be
addressed. First, as very nicely demonstrated in \cite{Dusling:2009df} that the microscopic interaction theory relevant for the medium can be related to
the momentum dependence of relaxation time, so taking a constant one undermines the momentum transfer of underlying interactions. Secondly, it gives 
rise to identical relaxation times for microscopic particle distributions and macroscopic fields like viscous flow, while the latter is expected to have 
a slower relaxation rate. Hence, in order to extract relativistic hydrodynamic theory, the application of momentum dependent relaxation time in transport 
equation is very indicative.

The current work demonstrates that the fundamental thermodynamic quantities such as energy density, pressure, particle number density and hydrodynamic
four velocity are redefined at each order of gradient expansion in an out of equilibrium situation including dissipative corrections from the medium. 
These corrections are estimated with momentum dependent medium interaction using MDRTA for the collision integral of the relativistic transport equation.
The transport coefficients for these corrections are estimated for the first and second order theory which turn out to be sensitively dependent upon the 
momentum dependence of medium interaction. Finally, the hydrodynamic evolution equations are obtained up to second order of gradient expansion.
The separation of relaxation time scales for microscopic particle distribution and macroscopic viscous fields have also been depicted which
again turns out to be crucially dependent upon the momentum dependence of medium interaction.

The manuscript is organised as follows. Section II provides a basic framework for hydrodynamic field redefinition obtained from relativistic
transport equation with a general collision integral. In section III the coefficients for first order field corrections are estimated using
MDRTA for the collision term in the transport equation. Section IV provides the second order corrections along with the coefficients and hydro
evolution equations. In section V the work has been summarized with necessary discussions and remarks.

\section{Field redefinition in relativistic hydrodynamic theory}

The formalism begins with the microscopic relativistic transport equation for the single particle distribution function $f(x,p)$ with particle four-momenta $p^{\mu}$ 
and space-time variable $x^{\mu}$,
\be
p^{\mu}\partial_{\mu}f(x,p)=C[f]=-\cal{L}[\phi]~.
\label{RTE}
\ee
$C[f]$ is the collision term which has been linearized to $\cal{L}[\phi]$ over the out of equilibrium distribution deviation $\phi$, ($f=f^{(0)}+f^{(0)}(1\pm f^{(0)})\phi$ with
$f^{(0)}$ as the equilibrium distribution) as follows,
\ba
{\cal{L}}[\phi]=&&\int d\Gamma_{p_1}d\Gamma_{p'}d\Gamma_{p'_1}f^{(0)}f_1^{(0)}(1\pm f'^{(0)})(1\pm f_{1}'^{(0)})\nn
&&\{\phi+\phi_1-\phi'-\phi'_1\}W(p'p'_1\mid pp_1)~.
\label{coll}
\ea
Here, $d\Gamma_p=\frac{d^3p}{(2\pi)^3p^0}$ is the phase space factor and $W$ is the interaction rate carrying microscopic cross section.
Before applying the MDRTA formalism, let us proceed with the general form of $\cal{L}[\phi]$ given in Eq.(\ref{coll}) for which I briefly review
few of its properties. Energy-momentum and particle number conservation gives ${\cal{L}}[p^{\mu}]=0$ and ${\cal{L}}[1]=0$ respectively. Self adjoint properties
$\int d\Gamma_p\psi{\cal{L}[\phi]}=\int d\Gamma_p\phi{\cal{L}[\psi]}$ with $\psi=\psi(x,p^{\mu})$ gives rise to summational invariant property 
$\int d\Gamma_p {\cal{L}[\phi]}=0$ and $\int d\Gamma_p p^{\mu} {\cal{L}[\phi]}=0$. Finally, applying non-negative entropy production rate it can be
proved that $\int d\Gamma_p\phi{\cal{L}[\phi]}\geq 0$ where the equality holds for $\phi=\{1,p^{\mu}\}$ \cite{Degroot}. 
If one proceeds to solve Eq.(\ref{RTE}) with order by order gradient expansion method, for each order $r$ ($\phi=\sum_r\phi^{(r)}$), the left hand side of (\ref{RTE})
turns out to be a linear combination of thermodynamic forces with different tensorial ranks as follows,
\ba
&&\sum_{l}Q_l^{(r)} X_l^{(r)}+\sum_m R_m^{(r)\mu}Y^{(r)}_{m\mu}+\sum_n S_n^{(r)\mu\nu}Z^{(r)}_{n\mu\nu}\nn
&&=-{\cal{L}}[\phi^{(r)}]~.
\label{TF1}
\ea
Here, $X_l^{(r)}, Y^{(r)}_{m\mu}$ and $Z^{(r)}_{n\mu\nu}$ respectively are the scalar, vector and rank-2 tensor thermodynamic forces of gradient expansion order $r$.
The indices $l,m$ and $n$ denote number of independent thermodynamic forces of each kind respectively for each order. For eg.,
for scalar forces with $r=1$, $l$ has only one value corresponding to $\partial\cdot u$. For $r=2$, $l$ runs over $D(\partial\cdot u), (\partial\cdot u)^2, 
\sigma_{\mu\nu}\sigma^{\mu\nu},\cdots$ and so on. From here on the macroscopic thermodynamic quantities are needed to be addressed.
$T,u^{\mu}$ and $\tilde{\mu}=\mu/T$ denote the temperature, hydrodynamic four-velocity and scaled chemical potential of the system respectively.
$D=u^{\mu}\partial_{\mu}$ and $\nabla^{\mu}=\Delta^{\mu\nu}\partial_{\nu}$ are temporal and spatial counterparts of the total space-time derivative 
$\partial^{\mu}=u^{\mu}D+\nabla^{\mu}$ with $\Delta^{\mu\nu}=g^{\mu\nu}-u^{\mu}u^{\nu}$ and $g^{\mu\nu}=(1,-1,-1,-1)$.
$\sigma_{\mu\nu}=\nabla_{\langle{\mu}}u_{\nu\rangle}$ with 
$\langle\rangle$ denoting the traceless irreducible tensors of rank-1 and 2 defined as $A_{\langle\mu\rangle}=\Delta_{\mu\nu}A^{\nu}$ 
and $ A_{\langle\mu} B_{\nu\rangle}  =\Delta_{\mu\nu\alpha\beta}A^{\alpha}B^{\beta}$ respectively with 
$\Delta_{\mu\nu\alpha\beta}=\frac{1}{2}\{\Delta_{\mu\alpha}\Delta_{\nu\beta}+\Delta_{\mu\beta}\Delta_{\nu\alpha}\}-\frac{1}{3}\Delta_{\mu\nu}\Delta_{\alpha\beta}$.
$Q_l^{(r)}, R_m^{(r)\mu}$ and $S_n^{(r)\mu\nu}$ are the contracted parts which carry the particle signature being function of $p^{\mu}$ and its 
scaled mass $z=m/T$. The general solution for $\phi^{(r)}$ is a linear combination of the thermodynamic forces as the following,
\ba
\phi^{(r)}=\sum_{l}A_l^{r} X_l^{(r)}+\sum_m B_m^{r\mu}Y^{(r)}_{m\mu}+\sum_n C_n^{r\mu\nu}Z^{(r)}_{n\mu\nu}~
\label{phi01}
\ea
where the unknown coefficients $A_l^r, B_m^{r\mu}$ and $C_n^{r\mu\nu}$ are needed to be estimated from the transport equation (\ref{RTE}) itself.  
Since the thermodynamic forces are independent it is straightforward to derive that,
\be
Q_l^{(r)}=-{\cal{L}}[A_l^r]~,R_l^{(r)\mu}=-{\cal{L}}[B_l^{r\mu}]~,S_l^{(r)\mu\nu}=-{\cal{L}}[C_l^{r\mu\nu}].
\label{phi02}
\ee
In order to extract the coefficients, here they are expanded in a simple polynomial basis following \cite{Degroot} as, 
\begin{align}
&A_l^r=\sum_{s=0}^p (A_l^r)^s(z,x) \tau_p^s,\\
&B_l^{r\mu}=B_l^r\tilde{p}^{\langle\mu\rangle}~,~~~B_l^r=\sum_{s=0}^p (B_l^r)^s(z,x) \tau_p^s,\\
&C_l^{r\mu\nu}=C_l^r\tilde{p}^{\langle\mu}\tilde{p}^{\nu\rangle}~,~~~C_l^r=\sum_{s=0}^p (C_l^r)^s(z,x) \tau_p^s,
\end{align}
where the series is expanded up to any desired degree of accuracy. Here $\tilde{p}^{\mu}=p^{\mu}/T$ is the scaled particle
four-momenta and $\tau_p=p^{\mu}u_{\mu}/T$ is scaled particle energy at local rest frame.
Following the previous mentioned properties of $\cal{L}[{\phi}]$, it can be observed that
$(A^r_l)^0,(A^r_l)^1$ and $(B^r_l)^0$ can not be determined by the transport equation and hence called the homogeneous solutions.
Beyond that, $A_l^r,B_l^r$ and $C_l^r$'s can be estimated from the microscopic transport equation and can be called interaction solutions. Keeping upto first nonvanishing contribution in thermodynamic
fluxes, the interaction solutions can be extracted as the following,
\begin{align}
&(A^r_l)^2=-\big\{\int d\Gamma_p\tau_p^2Q_l^r\big\}/[\tau_p^2,\tau_p^2]~,\\
&(B^r_l)^1=-\big\{\int d\Gamma_p\tau_p\tilde{p}^{\langle\nu\rangle}R_l^{r\mu}\big\}/[\tau_p\tilde{p}^{\langle\nu\rangle},\tau_p\tilde{p}^{\langle\mu\rangle}]~,\\ 
&(C^r_l)^0=-\big\{\int d\Gamma_p\tilde{p}^{\langle\alpha}\tilde{p}^{\beta\rangle}S_l^{r\mu\nu}\big\}/[\tilde{p}^{\langle\alpha}p^{\beta\rangle},\tilde{p}^{\langle\mu}p^{\nu\rangle}].
\end{align}
The bracket quantity is defined as, $[\phi,\phi]=\int d\Gamma_p\phi {\cal{L}}[\phi]$ which are always non-negative.

I next proceed to estimate the out of equilibrium corrections for particle number density, energy density and thermodynamic pressure with the help of 
Eq.(\ref{phi01}). These quantities are conventionally defined respectively as, $\rho=u^{\mu}N_{\mu}$, $\epsilon=u_{\mu}u_{\nu}T^{\mu\nu}$ 
and $P=-\frac{1}{3}\Delta_{\mu\nu}T^{\mu\nu}$ with the help of energy-momentum tensor $T^{\mu\nu}=\int d\Gamma_pp^{\mu}p^{\nu}f$ and particle 4-flow $N^{\mu}=\int d\Gamma_pp^{\mu}f$.
Following this prescription the respective corrections are $\delta \rho=\sum_r\delta {\rho}^{(r)}$, $\delta\epsilon=\sum_r\delta\epsilon^{(r)}$ and $\delta P=\sum_r\delta P^{(r)}$ 
where the corrections for each order are,  
\ba
&&\delta \rho^{(r)}=\sum_l (c_{\Gamma})^r_lX^{(r)}_l,~~\delta\epsilon^{(r)}=\sum_l (c_{\Lambda})^r_lX^{(r)}_l,\nn&&\delta P^{(r)}=\sum_l (c_{\Omega})^r_lX^{(r)}_l.
\label{del1}
\ea
The correction coefficients are given by,
\ba
(c_{\Gamma})^r_l&&=T\big\{(A^r_l)^0a_1+(A^r_l)^1a_2+(A^r_l)^2a_3\big\}~,
\label{coeffn}\\
(c_{\Lambda})^r_l&&=T^2\big\{(A^r_l)^0a_2+(A^r_l)^1a_3+(A^r_l)^2a_4\big\}~,
\label{coeffen}\\
(c_{\Omega})^r_l&&=\frac{T^2}{3}\big\{(A^r_l)^0(a_2-z^2a_0)+(A^r_l)^1(a_3-z^2a_1)\nn&&+(A^r_l)^2(a_4-z^2a_2)\big\}~.
\label{coeffP}
\ea
The corresponding vector corrections for mean-particle velocity ${\rho}_0\delta u^{\mu}_N=V^{\mu}=\sum_rV^{(r)\mu}$ and energy flow or momentum density 
$(\epsilon_0+P_0)\delta u_{E}^{\alpha}=W^{\alpha}=\sum_r W^{(r)\alpha}$ belonging to $r^{th}$ order  are respectively given by,
\ba
&W^{(r)\alpha}=\Delta^{\alpha}_{\mu}u_{\nu}\delta T^{(r)\mu\nu}=\sum_l(c_{\Sigma})^r_lY_l^{(r)\alpha}~,\\
&V^{(r)\alpha}=\Delta^{\alpha\nu}\delta N^{(r)}_{\nu}=\sum_l(c_{\Xi})^r_lY_l^{(r)\alpha}~,
\label{del2}
\ea
with,
\ba
(c_{\Sigma})^r_l&&=T^2\big\{(B^r_l)^0b_1+(B^r_l)^1b_2\big\}~,
\label{coeffW}\\
(c_{\Xi})^r_l&&=T\big\{(B^r_l)^0b_0+(B^r_l)^1b_1\big\}~,
\label{coeffV}
\ea
where $\delta N^{(r)\mu}$ and $\delta T^{(r)\mu\nu}$ are the $r^{th}$ order out of equilibrium correction to particle four-flow and energy-momentum tensor respectively.
The moment integrals
are given by,
\begin{align}
&a_n=\int dF_p \tau_p^n~\\, 
&\Delta^{\mu\nu}b_n=\int dF_p \tilde{p}^{\langle\mu\rangle}\tilde{p}^{\langle\nu\rangle}\tau_p^n~, \\
&\Delta^{\alpha\beta\mu\nu}c_n=\int dF_p\tilde{p}^{\langle\mu}\tilde{p}^{\nu\rangle}\tilde{p}^{\langle\alpha}\tilde{p}^{\beta\rangle}\tau_p^n~,
\end{align}
with $dF_p=d\Gamma_pf^{(0)}(1\pm f^{(0)})$.

Including these corrections the most general expressions for particle four-flow and energy-momentum tensor are given by,
\ba
N^{\mu}=&&({\rho}_0+\delta {\rho})u^{\mu}+V^{\mu}~,
\label{numberflow}\\
T^{\mu\nu}=&&(\epsilon_0+\delta\epsilon)u^{\mu}u^{\nu}-(P_0+\delta P)\Delta^{\mu\nu}\nn
&&+(W^{\mu}u^{\nu}+W^{\nu}u^{\mu})+\pi^{\mu\nu}~,
\label{enmomflow}
\ea
with $0$ subscript defining the respective equilibrium scalar quantities, $u^{\mu}$ being the equilibrium velocity and 
$\pi^{\mu\nu}=\Delta^{\mu\nu\alpha\beta}\delta T_{\alpha\beta}=\sum_r\sum_l(c_{\eta})^r_lZ_l^{r\mu\nu}$ 
as the shear stress tensor with $(c_{\eta})_l^r=T^2(C_l^r)^0c_0$.
Eq.(\ref{numberflow}) and (\ref{enmomflow}) along with (\ref{del1})-(\ref{coeffV}) set the out of equilibrium thermodynamic field definition of a system that includes dissipation.

Several issues need to be addressed here. First, in Eq.(\ref{coeffn},\ref{coeffen},\ref{coeffP},\ref{coeffW},\ref{coeffV}) only the interaction solutions $(A^r_l)^2,(B^r_l)^1$ (as well as $(C^r_l)^0$)
can be obtained from the transport equation (\ref{RTE}), the homogeneous solutions are fully arbitrary and can not be extracted from a microscopic theory.
Secondly, in (\ref{numberflow}) and (\ref{enmomflow}), the number of transport coefficients have been significantly increased as in the usual cases only the pressure
correction and any one of the vector fluxes do exist \cite{Romatschke:2009im,Denicol:2012es}. To show that these two issues are connected the two following
identities are obtained,
\ba
&&(c_{\Omega})^r_l-\big(\frac{\partial P_0}{\partial\epsilon_0}\big)_{{\rho}_0}(c_{\Lambda})^r_l-\big(\frac{\partial P_0}{\partial {\rho}_0}\big)_{\epsilon_0}(c_{\Gamma})^r_l=-(c_{\zeta})^r_l~,
\label{coeffreln1}\\
&&(c_{\Sigma})^r_l-\hat{h}T(c_{\Xi})^r_l=-\frac{T}{\hat{h}}(c_{\lambda})^r_l~,
\label{coeffreln2}
\ea
with $\hat{h}=(\epsilon_0+P_0)/{\rho}_0T$, $T,\tilde{\mu}$ belonging to the equilibrium value of $\epsilon$ and $\rho$.
Here, 
\ba
&&(c_{\zeta})^r_l=T^2\int dF_p\hat{Q}A_l^r~,\\
&&(c_{\lambda})^r_l=-\frac{1}{3}\frac{\epsilon_0+P_0}{n_0}\int dF_p\tilde{p}^{\mu}\tilde{p}_{\mu}(\tau_p-\hat{h})B_l^r~,
\ea 
are the $r^{th}$ order coefficients of bulk viscous flow $\Pi$ and diffusion flow $q^{\mu}$ respectively for $l^{th}$ kind of term with 
$\hat{Q}=\frac{z^2}{3}+\tau_p^2((\frac{\partial P_0}{\partial\epsilon_0})_{{\rho}_0}-\frac{1}{3})+\tau_p\frac{1}{T}(\frac{\partial P_0}{\partial {\rho}_0})_{\epsilon_0}$
\cite{Chakraborty:2010fr}. 
This leads to the fact that the out of equilibrium corrections to the thermodynamic quantities add up to produce
dissipative fluxes, 
\be\delta P-(\frac{\partial P_0}{\partial\epsilon_0})_{{\rho}_0}\delta\epsilon-(\frac{\partial P_0}{\partial {\rho}_0})_{\epsilon_0}\delta {\rho}=\Pi~,~~~ W^{\mu}-\hat{h}T V^{\mu}=q^{\mu},
\ee
with $\Pi=-\sum_r\sum_l(c_{\zeta})^r_l X^r_l$ and $q^{\mu}=-\sum_r\sum_l\frac{T}{\hat{h}}(c_{\lambda})^r_lY_l^{r\mu}$. 
Since, for $r=1$ it will be seen later that, $\hat{Q}f^{(0)}(1\pm f^{(0)})=\frac{1}{T}{\cal{L}}[A^1_1]$ and 
$\frac{1}{\hat{h}}\tilde{p}^{\langle\mu\rangle}(\tau_p-\hat{h})f^{(0)}(1\pm f^{(0)})=\frac{1}{T}{\cal{L}}[B_1^{1\mu}]$, applying collision operator properties it is found that 
$(c_{\zeta})^r_l$ and $(c_{\lambda})^r_l$ do not depend on the homogeneous solutions $(A_l^r)^0,(A_l^r)^1,(B_l^r)^0$ but uniquely specified by the interaction
solution $(A_l^r)^2$ and $(B_l^r)^1$. It can be trivially shown that, all the shear coefficients can be specified by $(C_l^r)^0$. 
The correction in thermodynamic quantities $\delta\epsilon,\delta \rho,\delta P, W^{\mu}$ and $V^{\mu}$ due to the arbitrary homogeneous part of $\phi$ is attributed solely to
the hydrodynamic frame choice recently extensively studied in \cite{Bhattacharya:2011tra,Bemfica:2017wps,Bemfica:2019knx,Kovtun:2019hdm,Kovtun:2012rj,Noronha:2021syv}.
From Eq.(\ref{coeffreln1}) and (\ref{coeffreln2}) it can be seen that the homogeneous part of the individual scalar and vector correction, i,e the frame information in thermodynamic quantities 
exactly cancels to retain only the interaction part in the transport coefficients of the dissipative fluxes at any order. This agrees with \cite{Kovtun:2019hdm} that not all transport
coefficients but their certain combinations remain invariant under field redefinition due to hydrodynamic frame choice. This part as already mentioned can not be
extracted from the microscopic dynamics of the system and remains arbitrary to certain choice. The interaction part of the dissipative correction
will be next estimated using the MDRTA technique from the relativistic transport equation. Frame choice and matching conditions
will be addressed in the later part of the work again.

\section{First order field corrections with MDRTA}

The relaxation time approximation is a simple method to linearize the collision term with the help of relaxation time $\tau_R$ of single particle
distribution function $f(x,p^{\mu})$ as follows, 
\be
\tilde{p}^{\mu}\partial_{\mu}f=-\frac{\tau_p}{\tau_R}f^{(0)}(1\pm f^{(0)})\phi ~,~~~~\tau_R(x,p)=\tau_R^0(x) \tau_p^n~,
\label{RTE1}
\ee
where the momentum dependence of $\tau_R$ is expressed as a power law of the scaled particle energy in a comoving frame with $\tau_R^0$ as the momentum independent part  
and $n$ as a number specifying the power of the scaled energy. 
In order to solve Eq.(\ref{RTE1}), here the well known iterative technique of gradient expansion, the Chapman-Enskog (CE) method has been adopted \cite{Degroot}. 
Following that, the first order correction to the particle distribution function is obtained as follows,
\be
\frac{\phi^{(1)}}{\tau_R^0}=
\tau_p^{n-1}\bigg[\hat{Q}\partial\cdot u+\big\{\frac{\tau_p}{\hat{h}}-1\big\}\tilde{p}^{\langle\mu\rangle}\nabla_{\mu}\tilde{\mu}+\tilde{p}^{\langle\mu}\tilde{p}^{\nu\rangle}\sigma_{\mu\nu}\bigg].
\label{phirelax}
\ee
In deriving Eq.(\ref{phirelax}), the equilibrium thermodynamic identities have been used such as $D{\rho}_0+{\rho}_0\partial\cdot u=0,D\epsilon_0+(\epsilon_0+P_0)\partial\cdot u=0$ 
and $(\epsilon_0+P_0)Du^{\mu}=\nabla^{\mu}P_0$ without the inclusion of any dissipative effects.

\begin{figure}[h]
\includegraphics[scale=0.35]{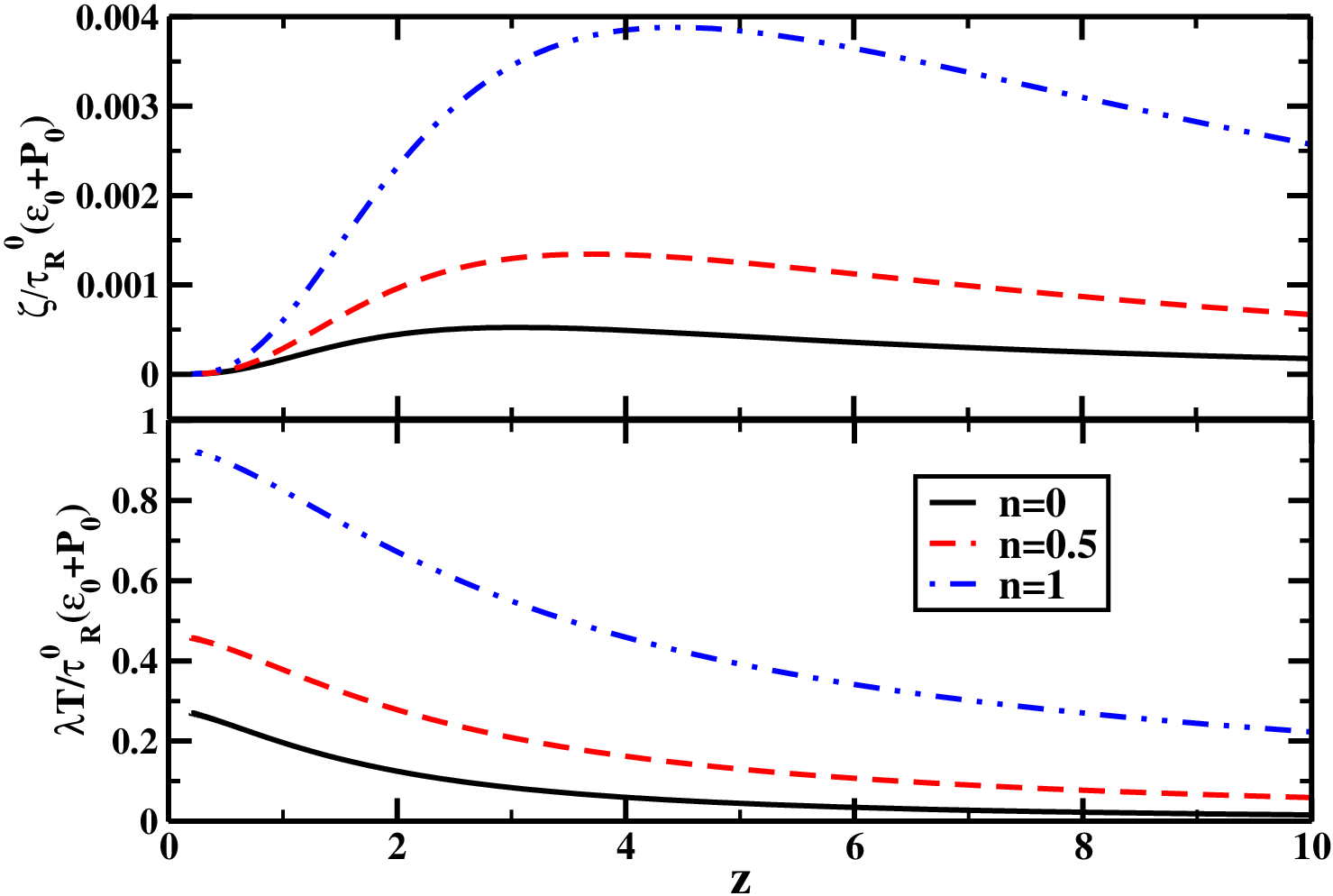} 
\caption{Scaled bulk viscosity and thermal conductivity as a function of $z$ with different $n$.}
\label{coeff}
\end{figure}

\begin{figure}[h]
\includegraphics[scale=0.35]{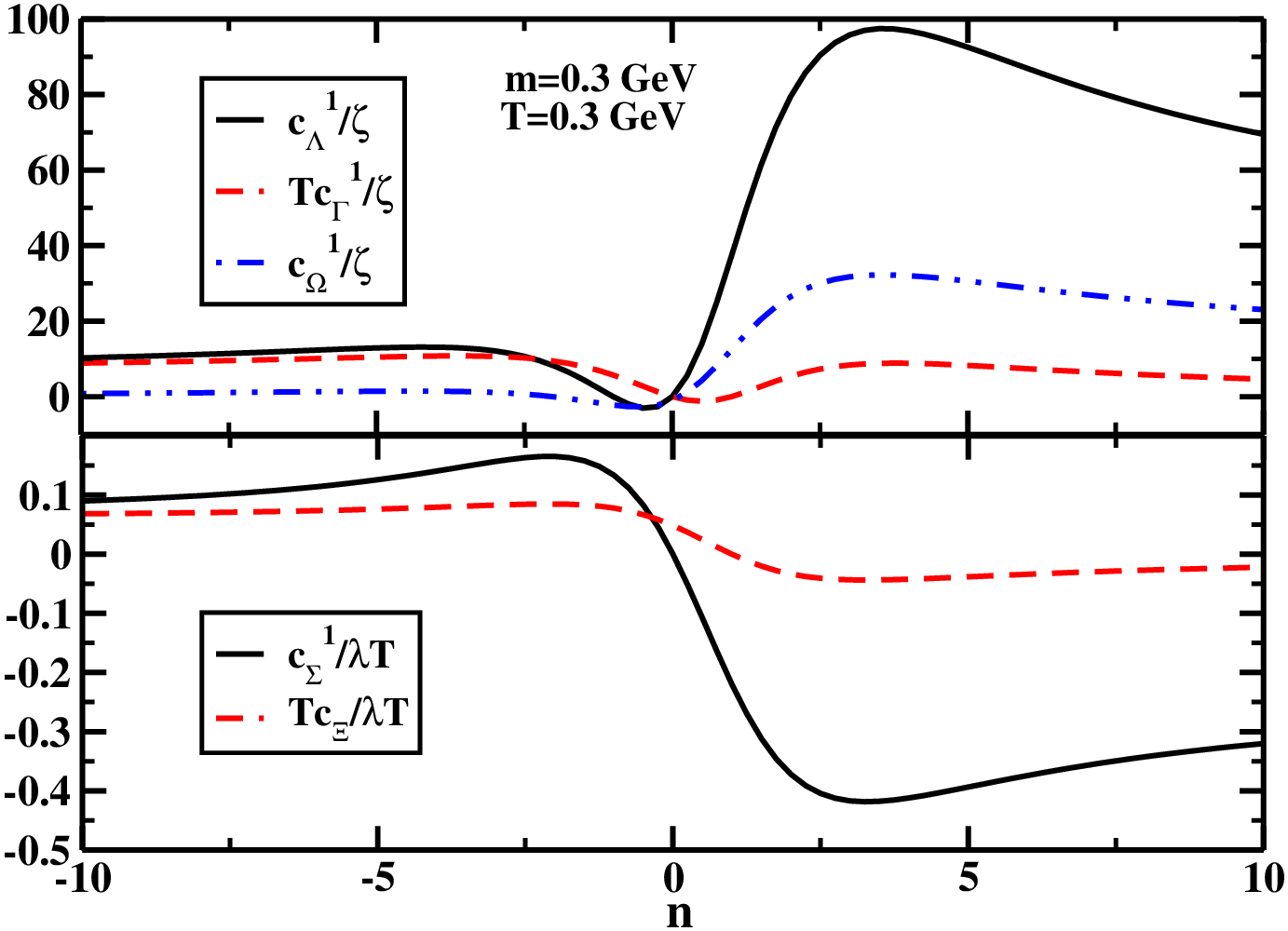} 
\caption{Correction coefficients as a function of $n$.}
\label{correction}
\end{figure}

Before proceeding further, the conservation of particle four-flow and energy-momentum tensor for $r=1$ needs to be checked. It can be proved,
\begin{align}
&\partial_{\mu}N^{\mu}=\int d\Gamma_pp^{\mu}\partial_{\mu}f=-\frac{T}{\tau_R^0}\int dF_p \tau_p^{1-n}\phi^{(1)}=0,
\label{cons1}\\
&\partial_{\mu}T^{\mu\nu}=\int d\Gamma_pp^{\nu}p^{\mu}\partial_{\mu}f\nn
&=-\frac{T^2}{\tau_R^0}\int dF_p\big\{u^{\nu}\tau_p^{2-n}+\tilde{p}^{\nu}\tau_p^{1-n}\big\}\phi^{(1)}=0,
\label{cons2}
\end{align}
for all values of $n$.
After achieving the conservation properties, the corresponding first order correction in thermodynamic quantities are given by,
\begin{align}
&\delta \epsilon^{(1)}=c_{\Lambda}^1(\partial\cdot u)~,~~\delta {\rho}^{(1)}=c_{\Gamma}^1(\partial\cdot u)~,~~\delta P^{(1)}=c_{\Omega}^1(\partial\cdot u)~,\nn
&W^{(1)\alpha}=-c_{\Sigma}\hat{h}(\nabla^{\alpha}T/T-Du^{\alpha})~,~~V^{(1)\alpha}=c_{\Xi}(\nabla^{\alpha}\tilde{\mu}),
\label{field1}
\end{align}
with associated correction coefficients,
\begin{align}
&\frac{c^1_{\Lambda}}{T^2\tau_R^0}=
\frac{z^2}{3}a_{n+1}+\bigg\{\big(\frac{\partial P_0}{\partial \epsilon_0}\big)_{{\rho}_0}-\frac{1}{3}\bigg\}a_{n+3}\nn
&~~~~~~~~+\frac{1}{T}\big(\frac{\partial P_0}{\partial {\rho}_0}\big)_{\epsilon_0} a_{n+2},
\label{coeff11}\\
&\frac{c^1_{\Gamma}}{T\tau_R^0}=
\frac{z^2}{3}a_{n}+\bigg\{\big(\frac{\partial P_0}{\partial \epsilon_0}\big)_{{\rho}_0}-\frac{1}{3}\bigg\}a_{n+2}+\frac{1}{T}\big(\frac{\partial P_0}{\partial {\rho}_0}\big)_{\epsilon_0} a_{n+1},
\label{coeff12}\\
&\frac{c^1_{\Omega}}{T^2\tau_R^0}=\frac{z^2}{9}a_{n+1}+\frac{1}{3}\bigg\{\big(\frac{\partial P_0}{\partial \epsilon_0}\big)_{{\rho}_0}-\frac{1}{3}\bigg\}a_{n+3}\nn
&~~~~~~~~+\frac{1}{3T}\big(\frac{\partial P_0}{\partial {\rho}_0}\big)_{\epsilon_0} a_{n+2}-\frac{z^4}{9}a_{n-1}\nn
&~~~~~~~~-\frac{z^2}{3}\big\{\big(\frac{\partial P_0}{\partial \epsilon_0}\big)_{{\rho}_0}-\frac{1}{3}\bigg\}a_{n+1}-\frac{z^2}{3T}\big(\frac{\partial P_0}{\partial {\rho}_0}\big)_{\epsilon_0} a_n,
\label{coeff13}\\
&\frac{c^1_{\Sigma}}{T^2\tau_R^0}=\frac{1}{\hat{h}}b_{n+1}-b_{n}~,
\label{coeff14}\\ 
&\frac{c^1_{\Xi}}{T\tau_R^0}=\frac{1}{\hat{h}}b_{n}-b_{n-1}~.
\label{coeff15}
\end{align}
It is to be noted here that for momentum independent case $n=0$, $c_{\Lambda}^1=c_{\Gamma}^1=c_{\Sigma}^1=0$. 
$c_{\Xi}^1$ vanishes for $n=1$ with $c_{\Lambda}^1=3c_{\Omega}^1$.
The $\tau_R^0$ in the denominator of the correction coefficients in (\ref{coeff11}-\ref{coeff15}) can be replaced by expressing it in terms of 
the independent transport coefficients associated with the dissipative fluxes of corresponding tensorial rank. 
Putting $\phi^{(1)}$ in the expression of first order dissipative fluxes namely bulk viscous flow, diffusion flow and shear viscous flow respectively,
\begin{align}
&\Pi^{(1)}=-T^2\int dF_p\hat{Q}\phi^{(1)}=-\zeta(\partial\cdot u)~,\\ 
&q^{(1)\alpha}=T^2\int dF_p\tilde{p}^{\langle\mu\rangle}(\tau_p-\hat{h})\phi^{(1)}=-\frac{\lambda T}{\hat{h}}\nabla^{\alpha}\tilde{\mu}~,\\
&\pi^{(1)\mu\nu}=T^2\int dF_p\tilde{p}^{\langle\alpha}\tilde{p}^{\nu\rangle}\phi^{(1)}=2\eta\sigma^{\mu\nu}~,
\end{align}
the corresponding first order transport coefficients bulk viscosity ($\zeta$), thermal conductivity ($\lambda$) and shear viscosity ($\eta$) in MDRTA
are given by,
\begin{align}
&\frac{\zeta}{T^2\tau_R^0}=\frac{z^4}{9}a_{n-1}+\bigg\{\big(\frac{\partial P_0}{\partial \epsilon_0}\big)_{{\rho}_0}-\frac{1}{3}\bigg\}^2a_{n+3}\nn
&+\frac{2z^2}{3T}\big(\frac{\partial P_0}{\partial {\rho}_0}\big)_{\epsilon_0} a_n+\frac{2}{T}\bigg\{\big(\frac{\partial P_0}{\partial \epsilon_0}\big)_{{\rho}_0}-
\frac{1}{3}\bigg\}\big(\frac{\partial P_0}{\partial {\rho}_0}\big)_{\epsilon_0} a_{n+2}\nn
&+\frac{1}{T^2}\big(\frac{\partial P_0}{\partial {\rho}_0}\big)^2_{\epsilon_0} a_{n+1}+\frac{2z^2}{3}\bigg\{\big(\frac{\partial P_0}{\partial \epsilon_0}\big)_{{\rho}_0}-\frac{1}{3}\bigg\}a_{n+1}~,
\label{zeta}\\
&\frac{\lambda T}{T^2\tau_R^0}=-\big\{b_{n+1}-2\hat{h}b_n+\hat{h}^2b_{n-1}\big\}~, 
\label{lambda}\\
&\frac{\eta}{T^2\tau_R^0}=\frac{1}{2}c_{n-1}~.
\label{eta}
\end{align}
It has been checked that $\zeta/\tau_R^0,\lambda/\tau_R^0,\eta/\tau_R^0>0$ for all values of $n$ for various combinations of $z,T$ and $\tilde{\mu}$.  
They have been plotted in Fig.(\ref{coeff}) as a function of $z$ for several $n$ values.
\\
\\
 \\
Shear viscosity has not been plotted since tensor contribution is not entering in the dissipative correction of any thermodynamic quantity. 
As predicted earlier the following two relations hold for any $n$ value,
\ba
&&c^1_{\Omega}-c_{\Lambda}^1\big(\frac{\partial P_0}{\partial\epsilon_0}\big)_{{\rho}_0}-c^1_{\Gamma}\big(\frac{\partial P_0}{\partial {\rho}_0}\big)_{\epsilon_0}=-\zeta~,
\label{coeffreln11}\\
&&c^1_{\Sigma}-\frac{(\epsilon_0+P_0)}{{\rho}_0}c^1_{\Xi}=-\frac{\lambda T}{\hat{h}}~,
\label{coeffreln22}
\ea
such that,
\begin{align}
&\delta P^{(1)}-(\frac{\partial P_0}{\partial\epsilon_0})_{{\rho}_0}\delta\epsilon^{(1)}-(\frac{\partial P_0}{\partial {\rho}_0})_{\epsilon_0}\delta {\rho}^{(1)}=\Pi^{(1)}~,\\
&W^{(1)\mu}-\hat{h}T V^{(1)\mu}=q^{(1)\mu}~.
\end{align}
Putting (\ref{field1}) and (\ref{coeff11}-\ref{coeff15}) into (\ref{numberflow}) and (\ref{enmomflow}), the particle four-flow and energy-momentum tensor is respectively obtained 
including the out of equilibrium dissipative effects in all the thermodynamic quantities up to first order of gradient expansion. Eq.(\ref{coeffreln11}) and (\ref{coeffreln22}) 
exhibit that the number of
independent transport coefficients are still the same as for the usual case ($\zeta,\lambda$ and $\eta$). 
So it is observed that MDRTA introduces non-equilibrium dissipative contributions in all the thermodynamic quantities essential to define $N^{\mu}$ and $T^{\mu\nu}$
through the exponent $n$ without changing the number of independent transport coefficients. For $n=0$ situation, i.e, without taking any momentum
dependence in collision integral, one returns to the usual scenario where energy correction, particle number correction and energy flux vanishes leaving the entire scalar dissipation to
pressure correction and vector dissipation to particle flux.
So it can be said that in a general situation the dissipative corrections in thermodynamic field variables is determined by how the medium interaction distributes
the respective dissipative fluxes among the scalar and vector fields.
The coefficients of first order dissipative correction in (\ref{coeff11}-\ref{coeff15}) (scaled by independent transport coefficients) have been plotted for m=0.3 GeV and 
T=0.3 GeV as a function of $n$ in Fig.(\ref{correction}). Fig.(\ref{correction}) shows that the individual field corrections take how much fractional
part of the dissipative flux, is decided by the value of $n$.
\\
\\

The entropy production $\partial_{\mu}S^{\mu}=-\int d\Gamma_p (ln f)p^{\mu}\partial_{\mu}f$ under first order MDRTA turns out to be,
\be
T\partial_{\mu}S^{\mu}=\zeta (\partial\cdot u)^2+2\eta\sigma^{\mu\nu}\sigma_{\mu\nu}+\frac{\lambda T}{\hat{h}^2}(\vec{\nabla}\tilde{\mu})^2~,
\ee
which is non-negative as long as $\zeta,\lambda,\eta\geq 0$. So the individual values of dissipative correction coefficients do not affect the positive entropy
production rate. In this context it needs to be mentioned that for some $n$ values the coefficients in Eq.(\ref{coeff11}-\ref{coeff15}) 
can take negative values as can be
seen from Fig.(\ref{correction}), but as 
argued in \cite{Monnai:2018rgs}, they are acceptable as long as $\zeta, \lambda, \eta$ and hence the entropy production is non-negative.

\section{Second order field corrections with MDRTA}
Before estimating the second order out of equilibrium correction to particle distribution function, one needs to estimate the equation of particle number density, energy density
and equation of motion including up to second order of gradient corrections.
Applying $\partial_{\mu}N^{\mu}=0$ and $\partial_{\mu}T^{\mu\nu}=0$,
the following thermodynamic identities are obtained,
\begin{align}
&D({\rho}_0+\delta {\rho}^{(1)})+({\rho}_0+\delta {\rho}^{(1)})(\partial\cdot u)\nn
&-V^{(1)\mu}Du_{\mu}+\nabla_{\mu} V^{(1)\mu}=0,\\
&D(\epsilon_0+\delta\epsilon^{(1)})+(\epsilon_0+P_0+\delta\epsilon^{(1)}+\delta P^{(1)})(\partial\cdot u)\nn
&-\pi^{(1)\mu\nu}\sigma_{\mu\nu}-2W^{(1)\mu}Du_{\mu}+\nabla_{\mu} W^{(1)\mu}=0~,\\
&(\epsilon_0+P_0+\delta\epsilon^{(1)}+\delta P^{(1)})Du^{\alpha}-\nabla^{\alpha}(P_0+\delta P^{(1)})\nn
&+\Delta^{\alpha}_{\nu}\nabla_{\mu}\pi^{(1)\mu\nu}-\pi^{(1)\alpha\mu}Du_{\mu}\nn
&+W^{(1)\alpha}(\partial\cdot u)+\Delta^{\alpha}_{\nu}DW^{(1)\nu}+W^{(1)\mu}\partial_{\mu}u^{\alpha}=0~.
\label{iden1}
\end{align}
Using these identities and applying the second order iteration of CE method in Eq.(\ref{RTE1}), the second order correction in particle distribution function is obtained
as follows,
\begin{widetext}
\begin{align}
&\frac{\phi^{(2)}}{\tau_R^0}=\nn
&-\frac{c_s^2}{\epsilon_0+P_0}\tau_p^{n+1}\bigg\{\pi^{(1)\mu\nu}\sigma_{\mu\nu}
-\Pi^{(1)}(\partial\cdot u)\bigg(1-c_{\Lambda}^1(c_s^2+1)+c_s^2T\partial_{T}c_{\Lambda}^{1}\bigg)+c_{\Lambda}^1 D\Pi^{(1)}\bigg\}\nn
&-\frac{1}{(\epsilon_0+P_0)}\tau_p^n \tilde{p}_{\mu}\bigg\{\nabla_{\nu}\pi^{(1)\mu\nu}-(1-c_s^2c_{\Lambda}^1)\nabla^{\mu}\Pi^{(1)}\bigg\}\nn
&-\frac{\tau_R^0}{2\eta}\tau_p^{2n-1}\tilde{p}^{\langle\mu}\tilde{p}^{\nu\rangle}D\pi^{(1)\mu\nu}-\frac{\tau_R^0}{2\eta}\tau_p^{2n-2}\tilde{p}^{\mu}\tilde{p}^{\nu}\tilde{p}^{\rho}\nabla_{\rho}\pi^{(1)}_{\mu\nu}
+\frac{\tau_R^0}{\zeta}\bigg\{\frac{z^2}{3}\tau_p^{2n-1}+(c_s^2-\frac{1}{3})\tau_p^{2n+1}\bigg\}D\Pi^{(1)}\nn
&+\frac{\tau_R^0}{\zeta}\bigg\{\frac{z^2}{3}\tau_p^{2n-2}+(c_s^2-\frac{1}{3})\tau_p^{2n}\bigg\}\tilde{p}^{\alpha}\nabla_{\alpha}\Pi^{(1)}
+\frac{\tau_R^0}{2\eta}\tilde{p}^{\langle\mu}\tilde{p}^{\nu\rangle}\tilde{p}^{\langle\alpha}\tilde{p}^{\beta\rangle}\pi^{(1)}_{\mu\nu}\sigma_{\alpha\beta}\bigg\{\tau_p^{2n-2}-(n-1)\tau_p^{2n-3}\bigg\}\nn
&+\frac{\tau_R^0}{2\eta}\tilde{p}^{\langle\mu}\tilde{p}^{\nu\rangle}\pi^{(1)}_{\mu\nu}(\partial\cdot u)\bigg[(c_s^2-\frac{1}{3})\tau_p^{2n}+\bigg\{\frac{n-1}{3}
-c_s^2(n+1)+c_s^2\frac{T\partial_T(\tau_R^0/2\eta)}{(\tau_R^0/2\eta)}\bigg\}\tau_p^{2n-1}+\frac{z^2}{3}\tau_p^{2n-2}-(n-1)\frac{z^2}{3}\tau_p^{2n-3}\bigg]\nn
&+\frac{\tau_R^0}{\zeta}\tilde{p}^{\langle\mu}\tilde{p}^{\nu\rangle}\sigma_{\mu\nu}\Pi^{(1)}\bigg[(\frac{1}{3}-c_s^2)\tau_p^{2n}-\frac{z^2}{3}\tau_p^{2n-2}
+(n+1)(c_s^2-\frac{1}{3})\tau_p^{2n-1}+\frac{z^2}{3}(n-1)\tau_p^{2n-3}\bigg]\nn
&+\frac{\tau_R^0}{\zeta}\Pi^{(1)}(\partial\cdot u)\bigg[-\frac{2}{3}z^2(c_s^2-\frac{1}{3})\tau_p^{2n}-\frac{z^4}{9}\tau_p^{2n-2}
-(c_s^2-\frac{1}{3})^2\tau_p^{2n+2}+(n-1)\frac{z^4}{9}\tau_p^{2n-3}\nn
&+\tau_p^{2n+1}\bigg\{-c_s^2(c_s^2-\frac{1}{3})T\frac{\partial_T(\tau_R^0/\zeta)}{(\tau_R^0/\zeta)}+(n+1)(c_s^2-\frac{1}{3})^2-c_s^2 T\partial_T c_s^2\bigg\}\nn
&+\tau_p^{2n-1}\bigg\{-c_s^2\frac{z^2}{3}T\frac{\partial_T(\tau_R^0/\zeta)}{(\tau_R^0/\zeta)}+\frac{2}{9}z^2+\frac{2}{3}(n+1)z^2(c_s^2-\frac{1}{3})\bigg\}\bigg]~.
\label{phi2}
\end{align}
\end{widetext}
The diffusive fluxes have been ignored in the above expression due to calculational complexity. $c_s^2=(\partial P_0)/(\partial \epsilon_0)_{{\rho}_0}$ is defined as the squared velocity of sound. 

Putting $\phi^{(2)}$ in Eq.(\ref{cons2}) one again obtains it to be zero, giving rise to $\partial_{\mu}T^{\mu\nu}=0$.
This demonstrates that dissipative corrections under MDRTA conserves the energy-momentum perfectly if the thermodynamic variables
are redefined properly in the previous order and that modification is incorporated in the thermodynamic identities accordingly.

\begin{figure}[h]
\includegraphics[scale=0.36]{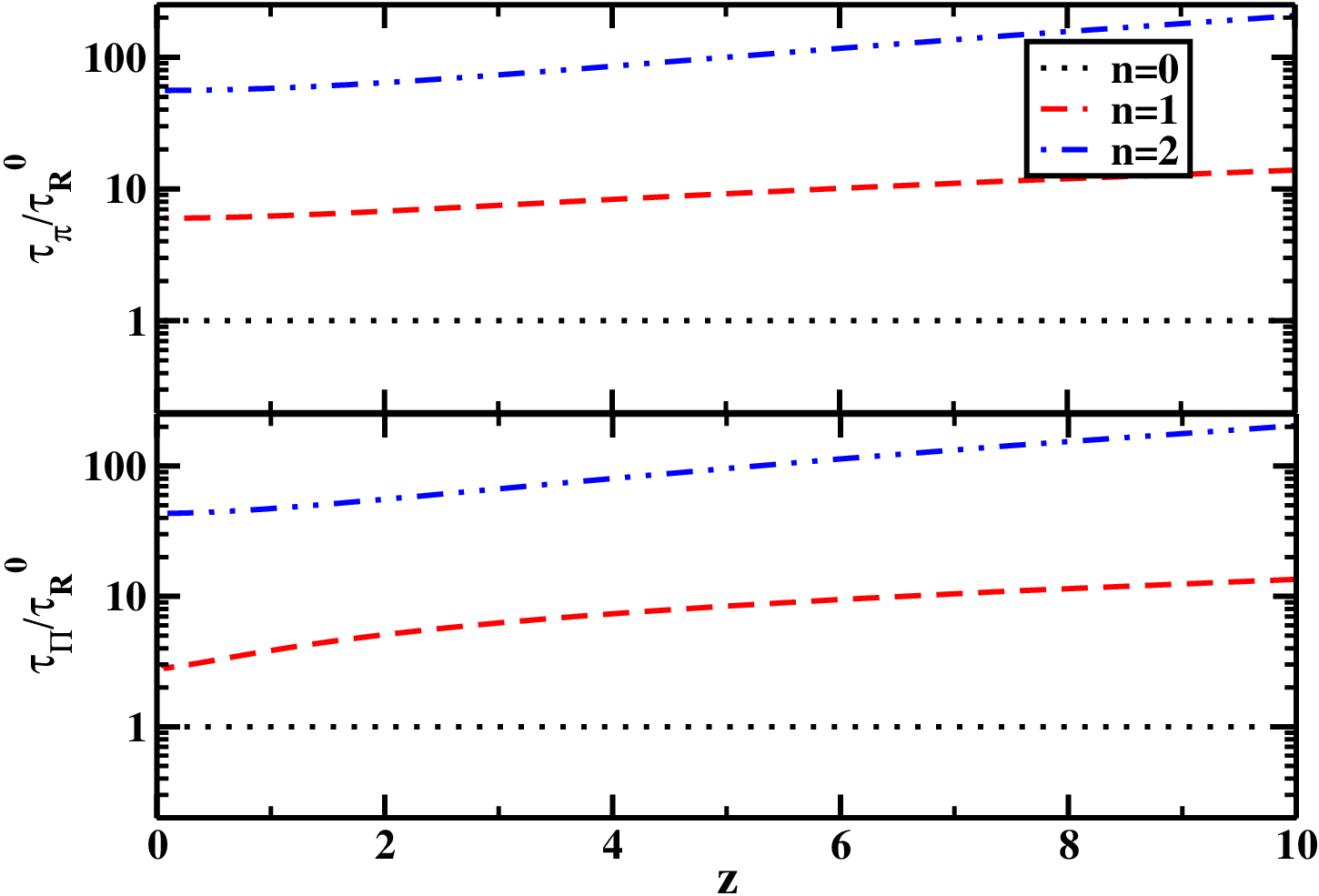} 
\caption{$\tau_{\pi}/\tau_R^0$ and $\tau_{\Pi}/\tau_R^0$ as a function of $z$ with different $n$.}
\label{relax}
\end{figure}

Utilising Eq.(\ref{phi2}), it is now customary to obtain the second order hydrodynamic equations for bulk and shear viscous flow by putting $\phi^{(2)}$
in $\Pi^{(2)}$ and $\pi^{(2)\mu\nu}$ respectively.
The bulk viscous and shear viscous pressure equations for a second order hydrodynamic theory with MDRTA is respectively given below along
with their transport coefficients.
\ba
&&\Pi=-\zeta\partial\cdot u -\tau_{\Pi}D\Pi+c_{\Pi}^{\sigma}\pi^{\mu\nu}\sigma_{\mu\nu}+c_{\Pi}^{\theta}\Pi(\partial\cdot u)~,
\label{bulkhydro2}\\
&&\pi^{\mu\nu}=2\eta\sigma^{\mu\nu}-\tau_{\pi}D\pi^{\langle\mu\nu\rangle}+c_{\pi}^{\omega}\pi_{\rho}^{\langle\mu}\omega^{\nu\rangle\rho}+c_{\pi}^{\sigma}\pi_{\rho}^{\langle\mu}\sigma^{\nu\rangle\rho}\nn
&&~~~~~~+c_{\pi}^{\theta}\pi^{\mu\nu}(\partial\cdot u)+c_{\pi}^{\zeta}\Pi\sigma^{\mu\nu}~,
\label{shearhydro2}
\ea
\begin{widetext}
\ba
&&\frac{\tau_{\Pi}}{\tau_R^0}= -\frac{T^2c_s^2}{(\epsilon_0+P_0)}c^1_{\Lambda}\bigg[\frac{z^2}{3}a_{n+1}+(c_s^2-1/3)a_{n+3}\bigg]  
+\frac{\frac{z^4}{9}a_{2n-1}+\frac{2}{3}z^2(c_s^2-\frac{1}{3})a_{2n+1}+(c^2_s-\frac{1}{3})^2a_{2n+3}}{\frac{z^4}{9}a_{n-1}+\frac{2}{3}z^2(c_s^2-\frac{1}{3})a_{n+1}+(c^2_s-\frac{1}{3})^2a_{n+3}}~,
\label{bulktau}\\
&& \frac{\tau_{\pi}}{\tau_R^0}=\frac{c_{2n-1}}{c_{n-1}}~,
\label{sheartau}\\
&&\frac{c_{\Pi}^{\sigma}}{\tau_R^0}=\frac{T^2c_s^2}{\epsilon_0+P_0}\bigg[\frac{z^2}{3}a_{n+1}+(c_s^2-\frac{1}{3})a_{n+3}\bigg]
+\bigg[\frac{2}{3}\bigg\{\frac{z^2}{3}a_{2n+1}+(c_s^2-\frac{1}{3})a_{2n+3}\bigg\}-\frac{2}{3}z^2\bigg\{\frac{z^2}{3}a_{2n-1}+(c_s^2-\frac{1}{3})a_{2n+1}\bigg\}\nn
&&~~~~~~~~~-\bigg\{\frac{z^2}{3}c_{2n-2}+(c_s^2-1/3)c_{2n}\bigg\}+(n-1)\bigg\{\frac{z^2}{3}c_{2n-3}+(c_s^2-1/3)c_{2n-1}\bigg\}\bigg]/c_{n-1}~,\\
&&\frac{c_{\Pi}^{\theta}}{\tau_R^0}=\frac{T^2c_s^2}{\epsilon_0+P_0}\big(-1+c_{\Lambda}^1(1+c_s^2)-c_s^2T\partial_Tc^1_{\Lambda}\big)\bigg\{\frac{z^2}{3}a_{n+1}+(c_s^2-1/3)a_{n+3}\bigg\}\nn
&&~~~~~~-\bigg[-\frac{2}{3}z^2(c_s^2-1/3)\bigg\{\frac{z^2}{3}a_{2n}+(c_s^2-1/3)a_{2n+2}\bigg\}-\frac{z^4}{9}\bigg\{\frac{z^2}{3}a_{2n-2}+(c_s^2-1/3)a_{2n}\bigg\}\nn
&&~~~~~~-(c_s^2-1/3)^2\bigg(\frac{z^2}{3}a_{2n+2}+(c_s^2-1/3)a_{2n+4}\bigg)+(n-1)\frac{z^4}{9}\big\{\frac{z^2}{3}a_{2n-3}+(c_s^2-1/3)a_{2n-1}\bigg\}\nn
&&~~~~~~+\bigg\{-c_s^2(c_s^2-1/3)\frac{T\partial_T(\tau_R^0/\zeta)}{\tau_R^0/\zeta}-c_s^2T\partial_Tc_s^2+(n+1)(c_s^2-1/3)^2\bigg\}\bigg(\frac{z^2}{3}a_{2n+1}+(c_s^2-1/3)a_{2n+3}\bigg)\nn
&&~~~~~~+\bigg\{-c_s^2\frac{z^2}{3}\frac{T\partial_T(\tau_R^0/\zeta)}{\tau_R^0/\zeta}+\frac{2}{9}z^2+\frac{2}{3}(n+1)z^2(c_s^2-\frac{1}{3})\bigg\}\bigg(\frac{z^2}{3}a_{2n-1}+(c_s^2-1/3)a_{2n+1}\bigg)\bigg]/\nn
&&~~~~~~~~~~\bigg[\frac{z^4}{9}a_{n-1}+\frac{2}{3}z^2(c_s^2-1/3)a_{n+1}+(c_s-1/3)^2a_{n+3}\bigg]~,\\
&&\frac{c_{\pi}^{\sigma}}{\tau_R^0}=\bigg[2c_{2n-1}-\frac{4}{7}\bigg\{c_{2n}-z^2c_{2n-2}-(n-1)c_{2n-1}+z^2(n-1)c_{2n-3}\bigg\}\bigg]/c_{n-1}~,~~~~~~~~c_{\pi}^{\omega}=2\tau_{\pi}~,\\
&&\frac{c_{\pi}^{\theta}}{\tau_R^0}=\bigg[(c_s^2-\frac{1}{3})c_{2n}+c_{2n-1}\bigg\{(n+1)(\frac{1}{3}-c_s^2)+c_s^2 \frac{T\partial_T(\tau_R^0/2\eta)}{(\tau_R^0/2\eta)}\bigg\}
+\frac{z^2}{3}c_{2n-2}-(n-1)\frac{z^2}{3}c_{2n-3}\bigg]/c_{n-1}~,\\
&&\frac{c_{\pi}^{\zeta}}{\tau_R^0}=\frac{\big[-(c_s^2-1/3)c_{2n}+(n+1)(c_s^2-1/3)c_{2n-1}-\frac{z^2}{3}c_{2n-2}+(n-1)\frac{z^2}{3}c_{2n-3}\big]}
 {\big[\frac{z^4}{9}a_{n-1}+\frac{2}{3}z^2(c_s^2-1/3)a_{n+1}+(c_s-1/3)^2a_{n+3}\big]}~.
\ea
\end{widetext}
The $\tau_R^0$ in the denominator can be related to $\zeta$ and $\eta$ respectively from (\ref{zeta}) and (\ref{eta}). 
Form Eq.(\ref{bulktau}) and (\ref{sheartau}) it can be observed that $\tau_{\pi}=\tau_{\Pi}=\tau_R^0$ holds only for $n=0$. For all other $n$, the three time scales are evidently separate. 
In Fig.(\ref{relax}) it has been shown that with increasing $n$ both $\tau_{\Pi}$ and $\tau_{\pi}$ become larger with respect to $\tau_R^0$ which is expected for the macroscopic
time scale. This separation of time scales with MDRTA itself provides a strong motivation for the study.
Analogous to the first order, the second order correction in thermodynamic quantities can be estimated as well. The second order energy and pressure corrections are respectively given by,
\ba
\delta\epsilon^{(2)}=c^2_{\Lambda} D\Pi^{(1)}+l^2_{\Lambda}\pi^{(1)\mu\nu}\sigma_{\mu\nu}+{\lambda}^2_{\Lambda}\Pi^{(1)}(\partial\cdot u)~,
\label{field2en}\\
\delta P^{(2)}=c^2_{\Omega} D\Pi^{(1)}+l^2_{\Omega}\pi^{(1)\mu\nu}\sigma_{\mu\nu}+{\lambda}^2_{\Omega}\Pi^{(1)}(\partial\cdot u)~.
\label{field2P}
\ea
For any $n$ value it can be shown that 
\be
c_{\Omega}^2-c_s^2 c^2_{\Lambda}=-\tau_{\Pi}~,~
l_{\Omega}^2-c_s^2 l^2_{\Lambda}=c_{\Pi}^{\sigma}~,~{\lambda}_{\Omega}^2-c_s^2 {\lambda}^2_{\Lambda}=c_{\Pi}^{\theta}~,
\ee
keeping the number of independent scalar transport coefficients same for all $n$ values.
This gives $\delta P^{(2)}-c_s^2\delta \epsilon^{(2)}=\Pi^{(2)}$. For $n=0,~ c^2_{\Lambda}, l^2_{\Lambda}, {\lambda}^2_{\Lambda}$ vanish as before and $\delta P^{(2)}$ becomes just $\Pi^{(2)}$.
The corresponding vector correction $W^{\mu}$ is, 
\be
W^{(2)\alpha}=r^2_{\pi}\Delta^{\alpha}_{\mu}\nabla_{\nu}\pi^{(1)\mu\nu}+r^2_{\Pi}\nabla^{\alpha}\Pi^{(1)}~,
\label{field2W}
\ee
with,
\ba
&&\frac{r^2_{\pi}}{\tau_R^0}=-\bigg[T^2\frac{b_{n+1}}{\epsilon_0+P_0}+\frac{c_{2n-1}}{c_{n-1}}\bigg]~,\\
&&\frac{r^2_{\Pi}}{\tau_R^0}=T^2(1-c_s^2c_{\Lambda}^1)\frac{b_{n+1}}{\epsilon_0+P_0}+\nn
&&\frac{\frac{z^2}{3}b_{2n-1}+(c_s^2-\frac{1}{3})b_{2n+1}}{\frac{z^4}{9}a_{n-1}+\frac{2}{3}z^2(c_s^2-\frac{1}{3})a_{n+1}+(c_s-\frac{1}{3})^2a_{n+3}}~,
\ea
both of which are zero for $n=0$. Note that $u^{\mu}\delta u_{\mu}=0$, which is essential for maintaining velocity normalization.  
Putting (\ref{field2en}),(\ref{field2P}) and (\ref{field2W}) in Eq.(\ref{enmomflow}), the second order $T^{\mu\nu}$ is obtained including
dissipative corrections using MDRTA.

\section{Summary and discussions}

In this work momentum dependent relaxation time approximation has been used to redefine the thermodynamic fields in order to include the out of equilibrium dissipative effects 
up to second order in gradient correction. The key finding is that these corrections are not independent but constrained to give the dissipative flux of same tensorial rank
where the associated coefficients are sensitive to the interaction. 
The derived equations can be applied
for hydrodynamic simulations since they can be uniquely solved and their phenomenological consequences can be significant as observed in \cite{Monnai:2018rgs}.

Here comes the question regarding frame choice and matching conditions. Frame choice is a vector condition that defines the out of equilibrium velocity flow,
i.e, putting constraints on $W^{\mu}$ or $V^{\mu}$.
The flow can be defined either by setting $W^{\mu}=0$ such that $T^{\mu\nu}u_{\nu}=(\epsilon_0+\delta\epsilon)u^{\mu}$ (Landau frame) or by $V^{\mu}=0$
such that $N^{\mu}=(\rho_0+\delta \rho)u^{\mu}$ (Eckart frame). Conventionally, in both the frames $\delta\epsilon$ and $\delta \rho$ are both set to zero in order 
to define the out of equilibrium temperature and chemical potential. But in a number of recent studies \cite{Monnai:2018rgs,Osada:2011gx,Osada:2012yp} 
it has been shown that in presence of dissipation $\epsilon$ and $\rho$ can have extended matching conditions including dissipative effects both for Landau
and Eckart frame retaining positive entropy production rate and causality and stability of the theory. The frame condition is not hampered by their
presence since the flow direction $u^{\mu}$ is not directly influenced by them. However, the constitutive equations for the dissipative currents
(\ref{bulkhydro2}) and (\ref{shearhydro2}) are frame independent as always \cite{Monnai:2010qp} since the associated coefficients do not include the homogeneous part
of the solution.

It is interesting to note here that though the conventional frame definition \cite{Bhattacharya:2011tra,Bemfica:2017wps,Bemfica:2019knx,Kovtun:2019hdm} does not include microscopic 
dynamics and entirely decided by macroscopic constraints (otherwise Eq.(\ref{coeffreln1}) and (\ref{coeffreln2}) will not be invariant under frame redefinition
since ($c_{\zeta})_l^r$ and $(c_{\lambda})_l^r$ are sensitive to interaction), \cite{Tsumura:2006hnr,Tsumura:2009vm} present a different perspective regarding hydrodynamic frame
choice. It suggests that instead of being arbitrary, the macroscopic frame vector can be related to the underlying microscopic theory via particle momenta and applies renormalization group method
to establish such a relation. Interestingly, this microscopic frame definition exactly agrees with the results obtained here. 
$n=0$ gives $u_{\mu}u_{\nu}\delta T^{\mu\nu}=0,u_{\mu}\delta N^{\mu}=0$ and $W^{\mu}=0$ 
which indicates the Landau frame. $n=1$ gives $u_{\mu}\delta N^{\mu}=0, V^{\mu}=0$ and $\delta T^{\mu}_{\mu}=0$ which are the constraints
for Eckart frame proposed by Stewart \cite{Stewart} and obtained by their analysis. In \cite{Tsumura:2007wu} a stable first order theory has been established in Eckart frame with this constraints
where not only $\delta P$, but also $\delta\epsilon$ includes contribution from the bulk flow. 
However, defining hydrodynamic frames in terms of underlying microscopic kinetic
theories is a debatable issue since in \cite{Tsumura:2012ss} it has been argued that the Landau-Lifshitz frame is the unique relativistic hydrodynamic frame, since from a
macroscopic perspective the frame vector should be independent of particle momenta.

Here, few aspects need to be clarified. First, the field redefinition derived here is purely dissipative correction taken care by the medium interaction and should
not be confused with that due to thermodynamic frame choice mentioned in \cite{Bemfica:2017wps,Bemfica:2019knx,Kovtun:2012rj,Kovtun:2019hdm}. The corrections
are expressed in terms of the dissipative forces which is why only the spatial gradients over fields are appearing in the corrections and not
the time derivatives as for the other case. Second, the conservation of energy momentum and particle number shown here is purely
macroscopic. Eq.(\ref{cons1}) and (\ref{cons2}) do not hold for any arbitrary $\phi$. If only the field corrections are properly implemented in
thermodynamic identities, the $\phi$ obtained from transport equation for next order satisfies those equations such that it can be said that the dissipative
field correction and conservation laws are compatible with each other. The only case where conservation holds at microscopic level (form of $\phi$ does not matter)
known to author is the new collision term proposed in \cite{Rocha:2021zcw} where the mentioned integrals are identically zero without the need of
extracting $\phi$ from order by order gradient expansion.

\end{document}